# Status of the Radial Schrodinger Equation

Anzor A.Khelashvili[1,2] and Teimuraz P. Nadareishvili[1]

[1] *Inst. of High Energy Physics, Iv. Javakhishvili Tbilisi State University, University Str. 9, 0109, Tbilisi, Georgia*

[2] *St.Andrea the First-called Georgian University of Patriarchy of Georgia, Chavchavadze Ave.53a, 0162, Tbilisi, Georgia.*

*Corresponding author. Phone: 011+995-98-54-47 ;  E-mail address: teimuraz.nadareishvili@tsu.ge;*

**Abstract:** We show that equation for radial wave function in its traditional form is compatible with the full Schrodinger equation if and only if a definite additional constraint required. This constraint has a boundary condition form at the origin. Some of consequences are also discussed.

**Keywords:** Full Schrodinger equation, radial equation, boundary condition, singular potentials.

### 1. Introduction.

It is well known that the radial Schrodinger equation

$$\frac{d^2 u(r)}{dr^2} - \frac{l(l+1)}{r^2} u(r) + 2m[E - V(r)]u(r) = 0 \qquad (1)$$

plays a central role in quantum mechanics due to frequent encounter with the spherically symmetric potentials. In turn, this equation is obtained from the full 3-dimensional Schrodinger equation

$$\Delta \psi(\vec{\mathbf{r}}) + 2m[E - V(r)]\psi(\vec{\mathbf{r}}) = 0 \qquad (2)$$

after the separation of variables in spherical coordinates [1, 2].

Recently considerable attention has been devoted to the problems of self-adjoint extension (SAE) for the inverse squared $r^{-2}$ behaved potentials in the radial Schrodinger equation [3]. These problems are interesting not only from academic standpoint, but also due to large number of physically significant quantum-mechanical problems that manifest in such a behavior.

Hamiltonians with inverse squared like potentials appear in many systems and they have sufficiently rich physical and mathematical structures. Starting from 60s of the previous century, singular potentials were the subject of intensive studies in connection with non-renormalizable field theoretic models. Exhaustive reviews dedicated to singular potentials for that era can be found in [4-6].

It turned out that there are no rigorous ways of deriving definite boundary condition for the radial wave function $u(r)$ from the radial equation itself at the origin $r = 0$ in case of singular potentials.

Many authors content themselves by consideration only a square integrability of radial wave function and do not pay attention to its behavior at the origin. Of course this is permissible mathematically and the strong theory of linear differential operators allows for such approach [7-9]. There appears so-called SAE physics [3], in the framework of which among physically reasonable solutions one encounters also many curious results, such as bound states in case of repulsive potential [10] and so on. We think that these highly unphysical results are caused by the fact that without suitable boundary condition at the origin a functional domain for radial Schrodinger Hamiltonian is not restricted correctly [11].

Below we show that, owing to the singular character of transformation leading to Eq. (1) from Eq. (2), there appears extra delta function term, which plays a role of point-like source, interacting to wave function. Surprisingly enough, this term has not been noted earlier. From the requirement of its absence definite constraint follows on the radial wave function at the origin. It has a form of boundary condition but indeed has more importance than boundary condition. This fact can have a great influence on the further considerations of the radial equation.

### 2. Rigorous derivation of radial equation.

Let us mention, that the transition from Cartesian to spherical coordinates is not unambiguous, because the Jacobian of this transformation $J = r^2 \sin\theta$ is singular at $r = 0$ and $\theta = n\pi (n = 0,1,2,...)$. Angular part is fixed by the requirement of continuity and uniqueness. This gives the unique spherical harmonics $Y_l^m(\theta,\varphi)$.

We also note in regards to radial variable that, although $\vec{r} = 0$ is an ordinary point in full Schrodinger equation, it is a point of singularity in the radial equation and thus, knowledge of specific boundary behavior is necessary.

We have to bear in mind that the radial Eq.(1) is not independent equation but is derived from full 3-dimensional Schrodinger equation (2) and as it is underlined in many classical books on quantum mechanics, the final radial equation must be compatible with the primary full Schrodinger equation. Unfortunately, in our opinion, this consideration has not been extended to any concrete results [2, 12]. Though several discussions of mostly "beat around the bush"[1] nature exist in the literature (see, e.g. book of R. Newton [13]), the conclusions from these studies are largely conservative and cautious. It seems that without deeper exploration of the idea of compatibility, some significant point will be missing.

Armed with this idea, let us now look at derivation of the radial wave equation in more details. Remembering that, after substitution

$$\psi(\vec{r}) = R(r) Y_l^m(\theta,\varphi) \tag{3}$$

into the 3-dimensional Equation (2), it follows the usual form of equation for full radial function $R(r)$:

---

[1] "beat about the bush" – is not very common expression, and therefore often causes some misunderstanding. This expression means: *approach a subject without coming to the point* (See, e.g. A.S.Hornby with A.P.Cowie "Oxford Advanced Lerner's Dictionary of Current English" special Edition for the USSR, Oxford Univ. Press, Oxford, 1982.

$$\frac{d^2R}{dr^2} + \frac{2}{r}\frac{dR}{dr} + 2m[E - V(r)]R - \frac{l(l+1)}{r^2}R = 0 \tag{4}$$

It is traditional trick to avoid the first derivative term from this equation by substitution

$$R(r) = \frac{u(r)}{r} \tag{5}$$

This substitution enhances singularity at $r = 0$, therefore we must be careful to perform it. Let us rewrite the equation (4) after this substitution

$$\frac{1}{r}\left(\frac{d^2}{dr^2} + \frac{2}{r}\frac{d}{dr}\right)u(r) + u(r)\left(\frac{d^2}{dr^2} + \frac{2}{r}\frac{d}{dr}\right)\left(\frac{1}{r}\right) + 2\frac{du}{dr}\frac{d}{dr}\left(\frac{1}{r}\right) - \left[\frac{l(l+1)}{r^2} - 2m(E - V(r))\right]\frac{u}{r} = 0 \tag{6}$$

We write equation in this form deliberately, indicating action of radial part of Laplacian on relevant factors explicitly. It seems that the first derivatives of $u(r)$ cancelled and we are faced to the following equation

$$\frac{1}{r}\left(\frac{d^2u}{dr^2}\right) + u\left(\frac{d^2}{dr^2} + \frac{2}{r}\frac{d}{dr}\right)\left(\frac{1}{r}\right) - \frac{l(l+1)}{r^2}\frac{u}{r} + 2m(E - V(r))\frac{u}{r} = 0 \tag{7}$$

Now if we differentiate the second term "naively", we'll derive zero. But it is true only in case, when $r \neq 0$. However in general this term is proportional to the 3-dimensional delta function. Indeed, taking into account that,

$$\frac{d^2}{dr^2} + \frac{2}{r}\frac{d}{dr} = \frac{1}{r^2}\frac{d}{dr}\left(r^2 \frac{d}{dr}\right) \equiv \Delta_r$$

is the radial part of the Laplace operator and [14]

$$\Delta_r\left(\frac{1}{r}\right) = \Delta\left(\frac{1}{r}\right) = -4\pi\delta^{(3)}(\vec{r}) \tag{8}$$

we obtain the equation for $u(r)$

$$\frac{1}{r}\left[-\frac{d^2u(r)}{dr^2} + \frac{l(l+1)}{r^2}u(r)\right] + 4\pi\delta^{(3)}(\vec{r})u(r) - 2m[E - V(r)]\frac{u(r)}{r} = 0 \tag{9}$$

We see that there appears the extra delta-function term, which must be eliminated. Note that when $r \neq 0$, this extra term vanishes owing to the property of the delta function and if, in this case, we multiply this equation on $r$, we obtain the ordinary radial equation (1).

However if $r = 0$, multiplication on $r$ is not permissible and this the extra delta-function term remains in Eq. (9). Therefore one has to investigate this term separately and find another way to avoid it.

The term with 3-dimensional delta-function must be comprehended as being integrated over $d^3r = r^2 dr \sin\theta d\theta d\varphi$. On the other hand [14]

$$\delta^{(3)}(\vec{r}) = \frac{1}{|J|}\delta(r)\delta(\theta)\delta(\varphi) \tag{10}$$

Taking into account all the above mentioned relations, one is convinced that extra term still survives, but now in the one-dimensional form

$$u(r)\delta^{(3)}(\vec{r}) \to u(r)\delta(r) \tag{11}$$

Its appearance as a point-like source breaks many fundamental principles of physics, which is not desirable. The only reasonable way to remove this term without modifying Laplace operator or including compensating delta function term in the potential $V(r)$, is imposing the requirement

$$u(0) = 0 \qquad (12)$$

(note, that multiplication of Eq. (9) on $r$ and then elimination this extra term owing the property $r\delta(r) = 0$ is not legitimated procedure, because effectively it is equivalent to multiplication on zero).

Therefore we conclude that the radial equation (1) for $u(r)$ is compatible with the full Schrodinger equation (2) if and only if the condition $u(0) = 0$ is fulfilled. *The radial equation (1) supplemented by the condition (12) is equivalent to the full Schrodinger equation (2).* We see that constraint equation has a form of boundary condition.

## 3. Conclusions and remarks

Some comments are in order here: equation for $R(r) = \dfrac{u(r)}{r}$ has its usual form (4). Derivation of boundary behavior from this equation is as problematic as for $u(r)$ from Eq. (1). Problem with delta function arises only in the course of elimination of the first derivative. Now, after the condition (12) is established, it follows that the full wave function $R(r)$ is less singular at the origin than $r^{-1}$. Though, this conclusion could be hasty because the transition to Eq. (1) for $R(r)$ is not necessary. It is also remarkable to note that the condition (12) is valid whether potential is regular or singular. It is only consequence of particular transformation of Laplacian. Different potentials can only determine the specific way of $u(r)$ tending to zero at the origin and the delta function arises in the reduction of the Laplace operator every time. All of these statements can easily be verified also by explicit integration of Eq. (9) over a small sphere with radius $a$ tending it to zero at the end of calculations.

It seems very curious that this fact was unnoticed up till now in spite of numerous discussions [2,5,6,12,13]. Now, that this (boundary) condition has been established, many problems can be solved by taking it into account. Remarkably, all the results obtained earlier for regular potentials with the boundary condition (12) remain unchanged. In the most textbooks on quantum mechanics $r \to 0$ behavior is obtained from Eq. (1) in case of regular potentials. But we have shown that this equation takes place only together with boundary condition (12). On the other hand, for *singular potentials* this condition will have far-reaching implications. Many authors neglected boundary condition entirely and were satisfied only by square integrability. But this treatment, after leakage into the forbidden regions and through a self-adjoint extension procedure, sometimes yields curious unphysical results. Below we consider some simple examples, showing the differences, which arise with and without above mentioned boundary condition:

(i) Regular potentials

$$\lim_{r \to 0} r^2 V(r) = 0 \qquad (13)$$

in this case, after substitution at the origin $u \sim r^a$, it follows from indicial equation, that $a(a-1) = l(l+1)$, which gives two solutions $u \underset{r \to 0}{\sim} c_1 r^{l+1} + c_2 r^{-l}$ (see, any textbooks on quantum mechanics). For non-zero $l$-s the second solution is not square integrable and is ignored usually. But for $l = 0$, many authors discuss (see, e.g. page 352 in [12]) how to deal with this solution, which is square integrable near the origin. According to our result, this solution must be ignored. Moreover, $u \underset{r \to 0}{\sim} const$ solution is forbidden, because there appears delta-function after its substitution into the full Schrodinger equation. Therefore, we must require $u(0) = 0$ for *any* $l$.

(ii) Transitive singular potentials

$$\lim_{r \to 0} r^2 V(r) = -V_0 = const \tag{14}$$

$V_0 > 0$ corresponds to the attraction, while $V_0 < 0$ - to repulsion.

In this case, the indicial equation takes form $a(a-1) = l(l+1) - 2mV_0$, which has two solutions: $a = \frac{1}{2} \pm \sqrt{\left(l + \frac{1}{2}\right)^2 - 2mV_0}$. Therefore

$$u \underset{r \to 0}{\sim} c_1 r^{\frac{1}{2}+P} + c_2 r^{\frac{1}{2}-P} \; ; \; P = \sqrt{\left(l + \frac{1}{2}\right)^2 - 2mV_0} \tag{15}$$

It seems, that both solutions are square integrable at origin as long as $0 \leq P < 1$. Exactly this range is studied in most papers. whereas according to our boundary condition we have the following restriction $0 \leq P < \frac{1}{2}$. The difference is essential. Indeed, the radial equation has form

$$u'' - \frac{P^2 - 1/4}{r^2} u = 2mEu \tag{16}$$

Depending on whether $P$ exceeds $1/2$ or not, the sign in front of the fraction changes and one can derive attraction in case of repulsive potential and vice versa. Boundary condition avoids this unphysical region $\frac{1}{2} \leq P < 1$.

Notice that the boundary condition at the origin was a subject of many fairly textbooks [15-17], as well as scientific articles [18,19]. The authors come to the condition $u(0) = 0$ by various ways starting from the radial equation (1). But their considerations are mainly restricted to the case of regular potentials. As regards of singular potentials there is no common view and people considered Dirichlet, which coincides to (12), or Neumann boundary conditions, as well as their generalization – Robin boundary condition [20]. We underline once again that derived constraint (12) is valid both for regular as well singular potentials.

Moreover our result above tells that the radial equation by itself is valid only in case $u(0) = 0$. Hence consideration based on radial equation is improper. It is evident that the deeper mathematical study of radial Hamiltonian is permissible, but without (12) constraint these investigations would have mathematical importance only and they have nothing common with physics, except $u(0) = 0$.

Lastly, we note that the same holds for radial reduction of the Klein-Gordon equation, because in three dimensions it has the following form

$$(-\Delta + m^2)\psi(\vec{r}) = [E - V(r)]^2 \psi(\vec{r}) \qquad (17)$$

and the reduction of variables in spherical coordinates will proceed to absolutely same direction as in Schrodinger equation.

**Acknowledgements**

We want to thank Profs. Sasha Kvinikhidze and Parmen Margvelashvili for valuable discussions. A.Khelashvili is indebted to thank Prof. Boris Arbuzov for reading the manuscript.